\documentclass[showpacs,showkeys,preprintnumbers,amsmath,amssymb,preprint,a4paper,pre]{revtex4}

\usepackage{amsmath,amssymb}
\usepackage{epsfig,graphicx,color}
\usepackage[applemac]{inputenc}

\begin{document}

\title[Induction in von K\'arm\'an flows driven by ferromagnetic impellers]{Induction in a von K\'arm\'an flow driven by ferromagnetic impellers}

\author{Gautier Verhille${}^1$, Nicolas Plihon${}^1$, Mickaël Bourgoin${}^2$, Philippe Odier${}^1$, Jean-Fran\c{c}ois Pinton${}^1$}
\address{(1): Laboratoire de Physique, CNRS \& \'Ecole Normale Sup\'erieure de Lyon, UMR5672, Universit\'e de Lyon, 46 all\'ee d'Italie, F69007 Lyon (France)\\
(2): Laboratoire des \'Ecoulements Géophysiques et Industriels, \\
CNRS/UJF/INPG UMR5519, BP53, F38041 Grenoble (France)}

\begin{abstract} 
We study magnetohydrodynamics in a von K\'arm\'an flow driven by the rotation of impellers made of material with varying electrical conductivity and magnetic permeability. Gallium is the working fluid and magnetic Reynolds numbers of order unity are achieved. We find that specific induction effects arise when the impeller's electric and magnetic characteristics  differ from that of the fluid. Implications in regards to the VKS dynamo are discussed.
\end{abstract} 


\maketitle

\section{Introduction}
The choice of electromagnetic boundary conditions in dynamo studies and its 
implication on the underlying magnetohydrodynamic processes has long been an 
open issue. In experiments, a finite flow of an electrically conducting fluid is generally considered, and the outer medium is ultimately electrically insulating (with electromagnetic properties of vacuum). Studies have considered layers with various 
properties regarding  electrical conductivity $\sigma_w$ and / or magnetic 
permeability $\mu_w$. Motivations for this have varied considerably. In many 
numerical simulations, it has been computationally convenient to assume an infinite 
value of $\mu_w$, yielding an attachment of magnetic field lines normal to the wall~
\cite{Kageyama:PP:95}.  In some simulations of the geodynamo, a thin wall 
approach~\cite{Muller:01} has been implemented in order to account for the rapid 
changes in electrical conductivity at the Core-Mantle Boundary~\cite
{Glatzmaier:PEPI:95,Glatzmaier:PhysD:96}.  Experiments have tend to study extensively changes with boundary conditions. The  motivations is that the critical magnetic Reynolds number value for the onset of dynamo action 
must be lowered as much as possible in order to lie within reach of a given 
experimental facility. For instance, in the pioneering experiments in Riga and 
Karlsruhe a volume of stagnant liquid sodium around the flow was found to be very 
favorable~\cite{Stefani:ZAMM:08,Verhille:SSR:09}. In the same spirit, dynamo action 
in the von K\'arm\'an sodium (VKS) experiment~\cite{Monchaux:PRL:07} has only been observed in situations where the fluid is driven by the motion of soft iron impellers. When replaced by non-magnetic stainless steel impellers, no dynamo was generated at the highest stirring achievable with the mechanical drives. In addition, the dominant VKS dynamo mode is essentially an axisymmetric dipole, in contradiction with predictions based on the mean flow structure~{\cite{Monchaux:PoF:09,Marie:EPJB:03}.  It is the motivation of this work to understand better the influence of the boundary conditions imposed by the ferromagnetic impellers, in particular whether they change significantly  the induction processes or simply lower the threshold of a dynamo that would be reached at higher rotation rates of non-magnetic impellers, if that was possible. To this end, we have performed extensive induction measurements in a gallium von K\'arm\'an flow driven by impellers made of stainless steel, copper or soft-iron. 

The experimental device is first described. Induction measurements are then presented and detailed in section~\ref{sec:exp}. A physical interpretation of these measurements is developed in section~\ref{sec:interp}.  In the last section, we present a discussion of the impact of these mechanisms on the VKS experiments.

\section{Experimental setup}
A von K\'arm\'an flow is produced by the coaxial rotation of two impellers inside a 
stainless steel cylindrical vessel filled with liquid gallium - \textit{c.f.} figure~\ref
{fig:setup}(a). The cylinder radius  $R$ is 97~mm and its length is 323~mm. The 
impellers have a radius equal to 82.5~mm and are fitted to a set of 8 straight blades 
with height $10$~mm. The impellers are separated by a distance $H =203$~mm. 
They are driven by two AC-motors  which provide a constant rotation rate in the 
interval $(F_1, F_2) \in [0.5, 25]$~Hz. 

In most of the cases, the flow is driven by symmetric counter-rotation of the impellers at $F_1=F_2=F$. It is structured in two rotating cells in front of each impeller, separated by a large azimuthal shear layer in the mid plane of the cylinder. Driving can also be achieved using the rotation of one single disk, the other being at rest; the flow then  consists in a single rotating cell. In both cases, the fluid is also ejected radially outward by the rotating(s) disk(s); this drives an axial flow toward them along the cylinder axis and a recirculation in the opposite direction along the cylinder lateral boundary. 

\begin{figure}[ht!]
\rightline{\includegraphics[width=0.85\columnwidth]{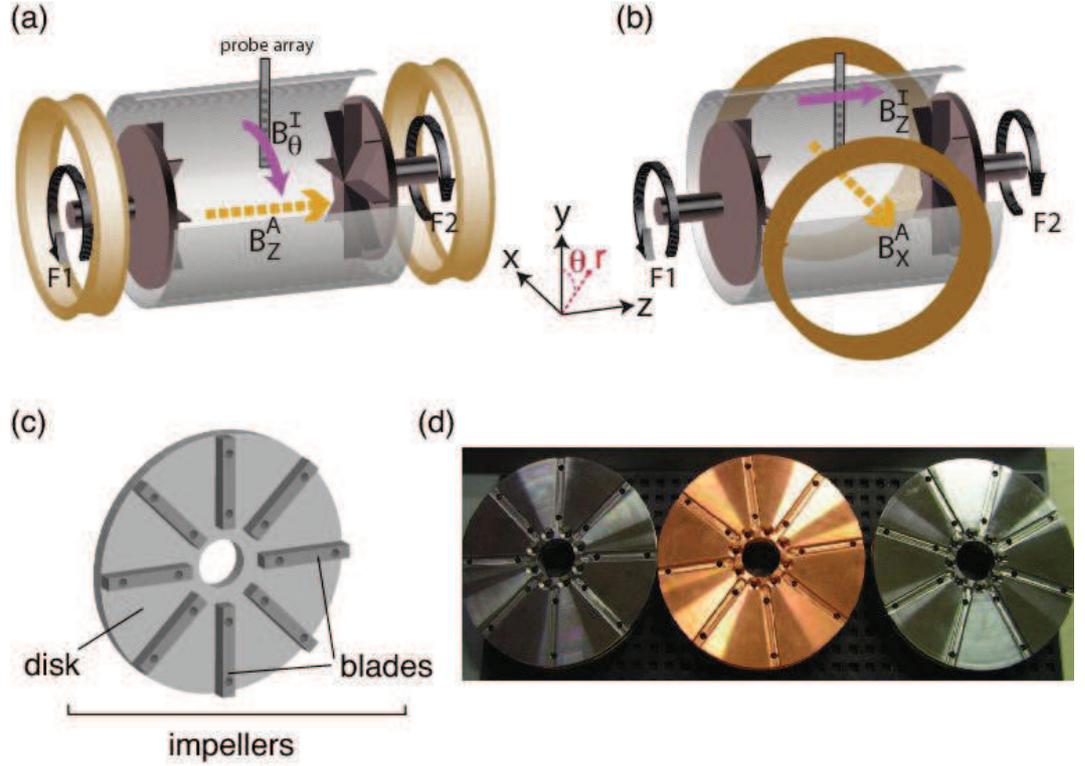}}
\caption{A von K\'arm\'an flow of liquid gallium is generated in a cylindrical vessel 
between two counter rotating impellers driven by two AC-motors. (a) When an axial 
field $B^A_z$ is applied, the counter rotating flow induces an azimuthal magnetic 
field  $B^I_\theta$ in the mid-plane of the flow.  A radial profile of this field is recorded  by a linear probe array of Hall sensors.  (b) When a transverse field $B^A_x$ is 
applied, an axial component $B^I_z$ is induced in the center, in a
plane perpendicular to the applied field. (c) Schematic and (d) pictures of the 
impellers used in this study: disks and blades can be assembled independently from 
elements made of stainless steel, copper or soft iron.}
\label{fig:setup}
\end{figure}

The system  is cooled by a water circulation located behind the driving 
impellers; the experiments are run at a temperature between $40^{\circ}$C and $48^
{\circ}$C. Liquid gallium  has density $\rho = 6.09 \times 10^{3} \; {\rm kg}\cdot{\rm m}^
{-3}$, electrical conductivity  $\sigma = 3.68 \times 10^{6} \; {\Omega}^{-1}\cdot{\rm m}
^{-1}$, hence a magnetic diffusivity $\lambda = 1/\mu_0\sigma = 0.22 \; {\rm m}^2\cdot {\rm s}^{-1}$. Its kinematic viscosity is  $\nu = 3.1 \times 10^{-7}  {\rm m}^{2}\cdot{\rm s} ^{-1}$.  The integral kinematic and magnetic Reynolds numbers are defined as ${\rm Re} = {2\pi R^{2}F}/{\nu}$ and ${\rm R_m} = 2\pi R^{2}F/\lambda$. By varying $F$, values of ${\rm R_m}$ up to $5$ are achieved, with corresponding ${\rm Re}$ values in excess of $10^6$.

Two pairs of induction coils (in a Helmoltz configuration) are aligned parallel to the rotation axis, or perpendicular to it. They create an applied magnetic field ${\mathbf B}^A$ of a few ten gauss inside the vessel, essentially uniform in the direction of the axis of the coils.  The interaction parameter\mbox{ $N=\sigma  (B^A)^{2} / 2 \pi  \rho  F \sim 10^{-5}$} is small, so that the action of Lorentz forces on the flow can be neglected. Magnetic induction measurements are performed using a Hall sensor probe array (8 probes) inserted into the flow in the mid-plane. Data are recorded using a National Instrument PXI-4472 digitizer at a rate of  $1000$~Hz with a 23 bits resolution.

The impellers driving the flow are made of a flat disk upon which straight blades are 
fixed. In the study reported here, both the disk and the blades can be chosen 
{independently} from stainless steel, copper or soft iron materials. These materials 
are characterized by their electrical conductivity and their magnetic permeability. For 
stainless steel and copper, the permeability is equal to that of vacuum, ${\mu_0=4\pi \times 10^{-7}}\ \text{H$\cdot$m}^{-1}$. The permeability of soft iron is known to change 
significantly during the manufacturing process. We have measured its value after 
machining from the $B-H$ response curves for small amplitude magnetic fields, using 
the procedure described in ~\cite{Oxley:JMMM:09}: a flux-meter is set on a soft iron 
blade and one on a copper blade, so as to build the hysteresis cycle of the soft iron 
$B=f(H)$. The permeability of soft iron is then computed from the slope $\partial B/
\partial H$ when $H \rightarrow 0$. The electrical and magnetic properties of the 
material used in this study are summarized in table~\ref{tab:prop}. Note that we 
checked experimentally and numerically -- using FEMM 4.2 software~\cite{FEMM42} 
--  that, with liquid gallium at rest and in the case of soft-iron impellers, the effect of the 
high permeability of the material produces only a weak distortion of the magnetic field 
applied by the coils at the probe location as compared to stainless steel impellers. 

\begin{table}
\hspace*{30mm}
\begin{tabular}{|l|c|c|c|c|} \hline\hline
                   & stainless steel & copper & soft iron	 & gallium\\ \hline\hline
$\mu_r$ 	&		1	  &	1	&	65	 &	1      \\ \hline
$\sigma$ [$\times10^6~\Omega^{-1}\cdot$m$^{-1}$] 	 & $\sigma_{ss}=1.4$ & 42 $
\sigma_{ss}$ & 7.3 $\sigma_{ss}$ & 2.6 $\sigma_{ss}$\\ \hline
$\lambda^{-1}=\mu\sigma$ [${\rm m}^{-2}\cdot {\rm s}$]  & $\lambda^{-1}_{ss}=1.8$ 	
& $42\ \lambda^{-1}_{ss}$ & $475\ \lambda^{-1}_{ss}$ & $2.6\ \lambda^{-1}_{ss}$\\ 
\hline\hline
\end{tabular}
\caption{Relative magnetic permeability ($\mu_r$), electric conductivity ($\sigma$) 
and magnetic diffusivity ($\lambda$) of the materials used, using stainless steel ($ss$) as a reference. Values taken from \cite
{Handbook:94}, except for iron permeability which was experimentally measured - 
see text.}\label{tab:prop}
\end{table}

\section{Induction linked to differential rotation}\label{sec:exp}
The swirling flow generated by the exact counter rotation of the impellers 
($F_1=F_2=F$) possesses a large shear layer in the mid-plane of the cylinder~\cite{Daviaud04a}.  Induction effects are generated by this azimuthal shear. We study the variation of their amplitude and radial profile, as the impellers are counter-rotated with increasing rotation rate $F$ and the impeller material varied. 

\subsection{Induction from an axial applied field}
We first investigate the case where the applied field is coaxial to the cylinder:  $
{\mathbf B^A} = B^A_z \hat{z}$. This configuration is sketched in figure~\ref{fig:setup}
(a). When stainless-steel impellers are used, the induced magnetic field along the azimuthal direction, denoted $B^I_\theta$, is due to the $\omega$-effect, and proportional to the fluid differential rotation $\partial_z v_\theta$~\cite{Odier:PRE:98,Bourgoin:MHD:04,Bourgoin:PoF:04}.

\begin{figure}[h!]
\rightline{\includegraphics[width=0.85\columnwidth]{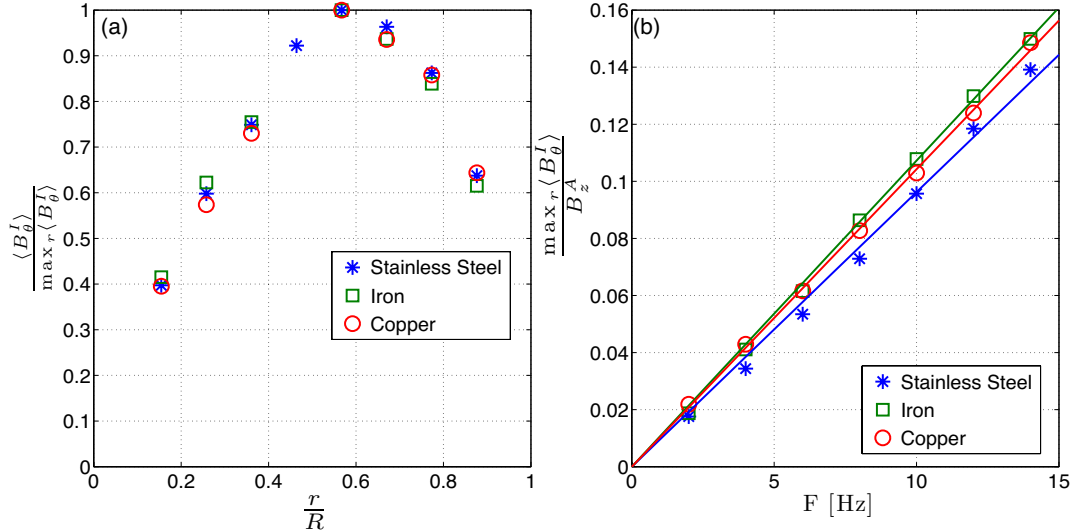}}
\caption{Time-averaged $\omega$-effect with an axially imposed field $B^A_z$. 
(a) Radial profiles of the induced field $\langle B^I_\theta\rangle$ at $F=12~$Hz for 
impellers made of stainless steel ($\star$), iron ($\square$) and copper ($\circ$). 
Profiles are normalized to the maximum of $\langle B^I_\theta\rangle$ along the 
radius ($\text{max}_r\langle B^I_\theta \rangle$). (b) Evolution of the maximum of $
\langle B^I_\theta\rangle$ along the radius normalized to the imposed field $B^A_z$ 
as a function of impellers rotation frequency (symbols as in (a)).}	
\label{fig:profil_axial}
\end{figure}

We now consider possible changes of this induction effect caused by driving the flow with impellers of varying materials.  The radial profiles (in the median plane $xOy$) of the time-averaged induced azimuthal field, $\langle B^I_\theta\rangle$,  are shown for $F$=12 Hz in figure~\ref{fig:profil_axial}.  Impellers made of stainless steel, copper or iron have been used (in this case, disks and blades are made of the same material). In figure~\ref{fig:profil_axial}(a), magnetic fields have been normalized to their maximum value in order to compare the induction {\it profiles}. We observed that they do not change noticeably when the material of the impellers is varied. The profiles are also independent of the impellers rotation rate. Figure~\ref{fig:profil_axial}(b) shows the evolution of the maximum amplitude of these radial profiles with the rotation rate $F$. One observes that the impeller material has a very weak influence here. Induction can be understood as the usual $\omega$-effect due to the  fluid motions (linear in the flow forcing and proportional to the time-
averaged fluid differential rotation $\langle \partial_z v_\theta\rangle$) with no noticeable influence of the impellers material. Note that all field amplitudes have been normalized to the applied field measured  at the probe location, thus taking into account the weak local distortion of the applied field lines in the case of the soft-iron impellers (the variation is weak -- less than 10~\%). \\

\begin{figure}[h!]
\rightline{\includegraphics[width=0.9\columnwidth]{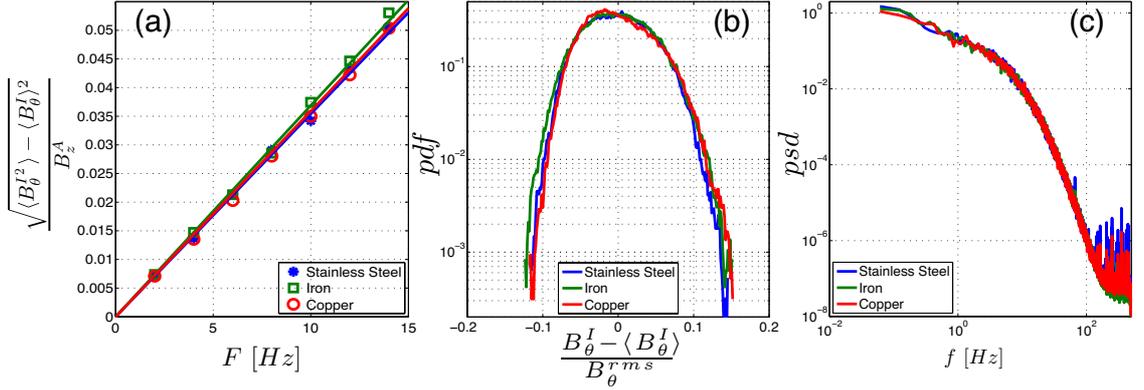}}
\caption{Fluctuations of the $\omega$-effect with an axially imposed field 
$B^A_z$. (a) Evolution of the standard deviation of $B^I_\theta$ as a function 
of the impellers rotation rate $F$ for impellers made of stainless steel ($\star$), iron ($
\square$) and copper ($\circ$). (b) Corresponding probability density function and (c) 
power spectrum density for $F = 12$~Hz and impellers made of stainless steel (blue), 
iron (green), copper (red).}
\label{fig:fluct_axial}
\end{figure}

Given the very high Reynolds number value of the flow, the induced magnetic field 
has a turbulent signature, with fluctuations as high as the time-averaged amplitude 
(see for instance~\cite{Odier:PRE:98}). The evolution of the $rms$ 
amplitude of the fluctuations as a function of the impellers rotation rate is displayed in 
figure~\ref{fig:fluct_axial}(a). It is linear in $F$ and the dispersion between data 
obtained for impellers made of the three materials is less than 5~\%. The probability 
density function (PDF) and power spectrum density (PSD) at $F = 12$~Hz are shown 
in figure~\ref{fig:fluct_axial}(b) and (c) respectively. These measurements show that 
no significant change in the fluctuations of the $\omega$-effect can be observed as the impellers material is varied -- at least at the location of the measurements, {\it i.e.}  here in the mid-plane.

\subsection{Induction from a transverse applied field}\label{sec:transverse}
In this section, we address the issue of induction from a transverse applied field: $
{\mathbf B^A} = -B^A_x \hat{x}$. This configuration is sketched in figure~\ref
{fig:setup}(b). The induced magnetic field is probed along the axial direction,  $B^I_z
$.

\begin{figure}[t!]
\rightline{\includegraphics[width=0.85\columnwidth]{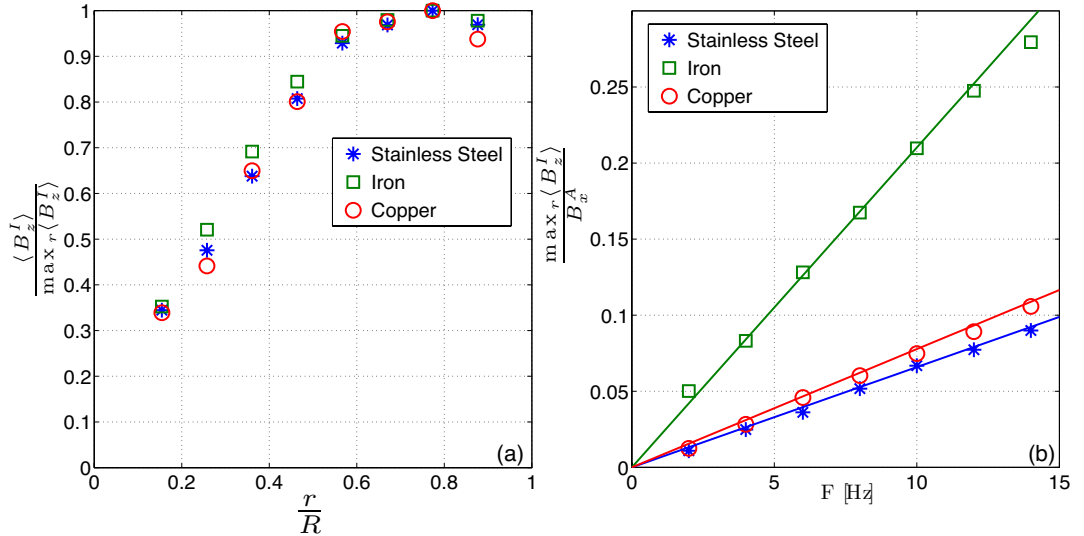}}
\caption{Time-averaged $BC$-effect with a transverse imposed field $B^A_x$. 
(a) Radial profiles of the induced field $\langle B^I_z\rangle$ at $F=12~$Hz for 
impellers made of stainless steel ($\star$), iron ($\square$) and copper ($\circ$). 
Profiles are normalized to the maximum of $\langle B^I_z\rangle$ along the radius ($
\text{max}_r\langle B^I_z \rangle$). (b) Evolution of the maximum of $\langle B^I_z
\rangle$ along the radius,  normalized to the imposed field $B^A_x$, as a function of 
impellers rotation frequency (symbols as in (a)).}
\label{fig:profil_transverse}
\end{figure}

For the case of a flow driven by stainless-steel impellers and enclosed in a stainless-steel vessel, the induction processes at work here have been described in details in~\cite{Bourgoin:MHD:04,Bourgoin:PoF:04} (in particular, see figure 13 in~\cite{Bourgoin:PoF:04}) and will only be recalled here. Near the impellers, the flow rotational motion advects the imposed magnetic field $B^A_x$, so as to induce a perpendicular component $B^I_y$. With the rotation opposite on either side of the flow mid-plane, two such contributions (with opposite directions) are generated on each side. These induced fields are associated to a current distribution consisting of two current sheets in the direction of the applied field, one in each flow cell, plus one with opposite direction in the shear layer. Since the outside medium is electrically
insulating, these currents loop back inside the flow volume. In doing so, they 
generate an axial field $B^I_z$, maximum in a plane transverse to the applied field. 
This effect, directly linked  to the rotation of the fluid, is linear in $F$. As in~\cite
{Bourgoin:PoF:04} we will refer to this induction process as a $BC$-effect in the 
what follows, since it originates in the Boundary Conditions from the steep variation in electrical conductivity at the flow wall. A simplified way to understand this effect is to consider the $B^I_z$ component in the mid-plane as resulting from the loop-back path of the $B^I_y$ contributions on either side. 

As before, we study modifications of this induction process resulting from changing the impellers. For the three materials used, the radial profile of the normalized time-averaged induced axial field $\langle B^I_z \rangle$ is shown in figure~\ref{fig:profil_transverse}(a). These normalized profiles are again independent of impellers rotation rate and impellers materials. Since the {\it profiles} are unchanged, one may suppose that a similar $BC$ mechanism can be invoked in all cases. In addition, in the low $R_m$ regimes accessible in the device, the $BC$-effect remains linear with $R_m$ whatever the impellers material. However, as seen in figure~\ref{fig:profil_transverse}(b), the {\it amplitude} of the time-averaged induced field now strongly depends on the impellers materials: a 20\% increase for copper and a 220\% increase for soft iron are measured, as compared to stainless steel impellers. \\

\begin{figure}[t!]
\rightline{\includegraphics[width=0.85\columnwidth]{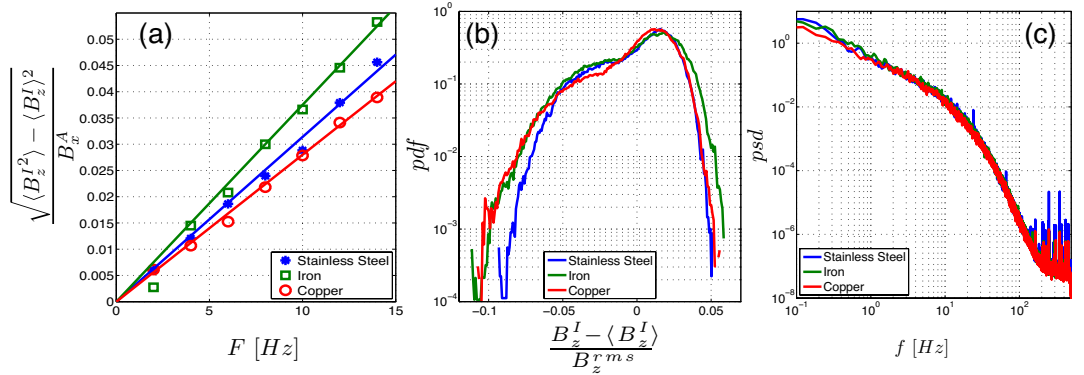}}
\caption{Fluctuations of the $BC$-effect with a transverse imposed field $B^A_x
$. (a) Evolution of the standard deviation of $B^I_z$ as a function of the 
impellers rotation rate $F$ for impellers made of stainless steel ($\star$), iron ($
\square$) and copper ($\circ$). (b) Corresponding PDF and (c) PSD at $F = 12$ Hz 
for impellers made of stainless steel (blue), iron (green), copper (red).}
\label{fig:fluct_trans}
\end{figure}

As shown in figure~\ref{fig:fluct_trans}, changes in the $rms$ intensity of the 
fluctuations of induction are much more modest. The variations are very small 
between stainless-steel and copper impellers. Soft-iron generate a 10\% increase of 
the fluctuation level compared to the other, non-magnetic, materials. This may be a second order effect: as the mean axial component $\langle B^I_z \rangle$ is increased, its stretching by fluctuations of the axial velocity gradient may add to the $rms$ intensity. Time spectra are however identical in the frequency range usually attributed to turbulent motions -- figure~\ref{fig:fluct_trans}(c). The bimodal distribution in figure~\ref{fig:fluct_trans}(b) is known to originate in the non-stationarity of the shear layer in the mid-plane~\cite{ravelet2004,Volk:PoF:06}. Its shape remains fairly independent of the impellers nature. Fluctuations of the induced field are thus not essentially modified by a change of impellers material. This may be expected since the magnetic field fluctuations trace back to the flow turbulence which is not expected to be affected by impellers material. 

We argue in section~\ref{sec:interp} below that the increase in the mean induced component $\langle B^I_z\rangle$ arises in a process similar to the one described in~\cite{Bourgoin:PoF:04}  from a localized $\omega$-effect generated when the magnetic diffusivity of the impellers differs from that of the fluid.

\section{Interpretation}\label{sec:interp}
\subsection{An induction mechanism associated to a jump of magnetic diffusivity at the impellers}
When probing induction effects in the mi- plane, our observations so far can be summarized as follows:\\
- in the case of an axial applied field (induction from differential rotation in the 
fluid bulk), the induced field seems to be fairly independent of the impeller material.\\
- in the case of a transverse applied field (induction from differential rotation, 
coupled with insulating boundary conditions), the time-averaged amplitude of the 
induced field varies strongly with  impeller material,\\
- in both cases, the spatial profile of the induced field is independent of the 
impellers nature, and the fluctuations of induction are not appreciably changed.\\

In order to interpret these observations, we need to understand the role of variations of magnetic diffusivity at the interface gallium-impellers on induction effects. We begin with the case of the transverse applied field, which shows the strongest dependence with the impellers nature.  Herzenberg and Lowes have studied the case of a finite cylinder rotating at frequency $\Omega$ in a uniform field perpendicular to the cylinder axis. To leading order equation 4.12 of~\cite{Herzenberg:PTRS:57} for the induced magnetic field writes:
\begin{equation}
{\mathbf B}^I(r, \theta, \phi) = \frac{1}{32\pi^2} (2\pi\mu\sigma \Omega a^2) B^A V_{\rm cyl} \hat{z} \times \nabla\left( \frac{\sin\theta\cos\phi}{ r^2}\right)  \ ,
\end{equation}
where $(r, \theta, \phi)$ are spherical coordinates with origin at the center of the cylinder of radius $a$ and $V_{\rm cyl}$ the volume of the cylinder. In the above derivation the cylinder and medium are assumed to have the same magnetic diffusivity, and the medium outside the cylinder is at rest. This induced field has approximately the geometry of a dipole with its axis perpendicular both to the axis of rotation and to the applied field. Its amplitude can also be written as 
$ B^I \propto B^A R_m^{\rm cyl} (V_{\rm cyl}/r^3)$, with $R_m^{\rm cyl} = \mu\sigma(2\pi a \Omega)a$ the magnetic Reynolds number of the rotating cylinder. It corresponds to a (radial) $\omega$-effect, due to the shearing of the magnetic field lines which are `advected' by the cylinder rotation; it is also the first step of the expulsion of the transverse field from the rotating body~\cite{Weiss:1966}. If we now consider a situation where the cylinder is made of a material with a different magnetic diffusivity, a similar induced field will be generated, even if the fluid around the cylinder rotates with the same velocity, {\it i.e.} if there is no differential rotation. This is because a change in magnetic permeability or a change in electrical conductivity also contribute to the induction processes.  For instance if the cylinder has a higher permeability, then the higher values of the flux of ${\mathbf B}$ in the cylinder will generate a shear of field lines at the fluid boundary. If the cylinder has a better electrical conductivity, then the same flux variation will be able to generate more induced currents. Altogether it is a change in the magnetic diffusivity $\mu\sigma$ that matters here. 

These arguments, supported by recent numerical works~\cite{PrivateNoreGisecke}, have lead us to write the correction to homogenous induction effects, in a first approximation, as
\begin{equation}
B^I \propto B^A (R_m^{\rm cyl} - R_m^{\rm fluid}) \frac{V_{\rm cyl}}{r^3} \ .
\end{equation}
for the induction generated by a local jump of either magnetic diffusivity or velocity at the cylinder end. \\ 

We then return to our induction measurements, as reported in the previous section. In the case of an applied field transverse to the flow axis, the rotation of each impeller creates an induced perpendicular dipole (as above). As in the $BC$-effect, the reconnection of these induced fields and the constraints on currents paths create an axial component in the $(yOz)$ plane. This field will add to the axial field induced by the usual $BC$ mechanism.  This explanation accounts for the very similar induction profile observed (see figure~\ref{fig:profil_transverse}(a)) with different materials used for the impellers: the additional effect due to the impellers is produced by a mechanism with a geometry identical to that of the $BC$-effect. It also explains why the impeller material have so little effect on the fluctuations of induction: the source of induction lies in the rotation of the impellers which is precisely controlled in the experiment to be constant in time, fixed to a prescribed value. Fluctuations are essentially due to changes in the induction generated by the velocity gradients within the flow, which have a quite non-stationary (turbulent) dynamics~\cite{Volk:PoF:06}. 

We thus write the axial induced field measured in the mid-plane as being the sum of two contributions: (i) the regular $BC$-effect which originates from the differential rotation of the fluid and (ii) the effect described above, caused by the change in magnetic diffusivity at the impellers: $B_z^I =  B_z^I({\rm fluid}) + B_z^I({\rm at \; impellers})$, {\it i.e.}
\begin{equation}
\frac{B_z^I}{B^A}=K_{\rm f}R_m + K_{\rm i} R_m \left(\frac{(\mu\sigma)_{\rm i}}{(\mu\sigma)_
{\rm f}}-1\right)\frac{V_{\rm i}}{V_{\rm i}^{\rm max}}
\label{eq:sum2}
\end{equation}
where $K_{\rm f, i}$ are constants taking into account the precise location of the measurement, and the second term has been chosen to emphasize two features: (i) no additional effect occurs when the impellers have a magnetic diffusivity which matches that of the fluid (all is then incorporated in the first term); (ii) the ratio $V_{\rm i}/V_{\rm i}^{\rm max}$ accounts for the fact that at each impeller, disk and blades can be made of materials; $V{\rm i}^{\rm max}$ is the maximum volume of disk plus blades, and $V_{\rm i}$ is the actual volume of a given material in the impeller (which can be disk and/or blade) so that the effective volume fraction is less than one -- $V_{\rm i} < V_{\rm i}^{\rm max} = 3.15~10^{-4} \; {\rm m}^3$.

\begin{figure}[h!]
\begin{center}
\includegraphics[width=0.7\columnwidth]{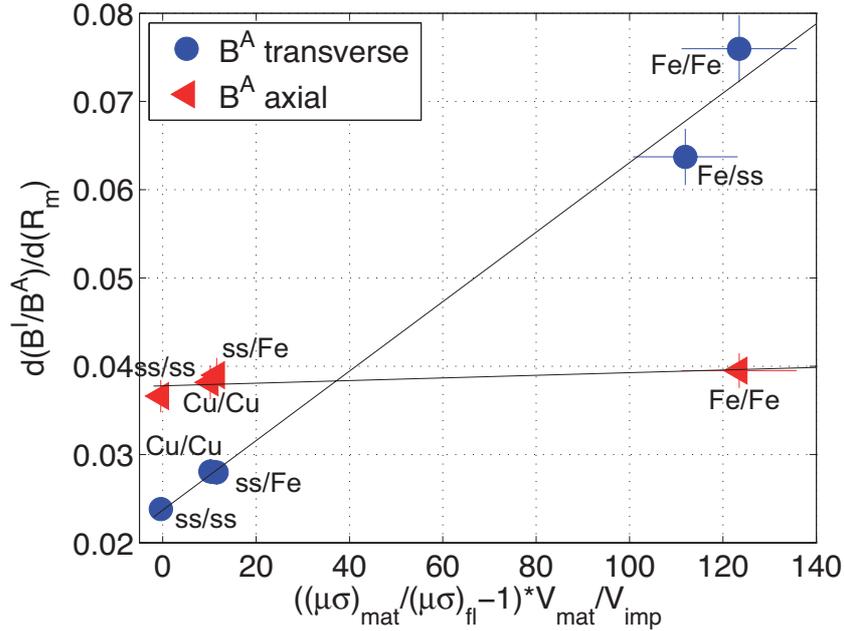}
\caption{Slopes of curves $B^I/B^A$ as a function of $R_m$, for  the different configurations studied, as a function of $\left(\frac{(\mu\sigma)_{\rm i}}{(\mu\sigma)_{\rm f}}-1\right)\frac{V_{\rm i}}{V_{\rm i}^{\rm max}}$ when the material of the disk and that of the blades are changed. The combinations used are indicated in the figure, with the first mention corresponding to the disc and the second to the blades. ``SS" stands for stainless steel, ``Fe" for iron and ``Cu" for copper. The blue circles correspond to the case of a transverse applied field and the red triangles to 
the axial applied field.}
\label{fig:slopes}
\end{center}
\end{figure}

By varying separately blade and disk materials, and performing measurements for a range of rotation rates of the impellers, one may thus probe the above relationship. Figure~\ref{fig:slopes} (blue circles) shows the slopes ${\frac{{\rm d}(B_z^I/B^A)}{{\rm d} R_m}}$ as a  function of the impeller parameter  $\left(\frac{(\mu\sigma)_{\rm i}}{(\mu\sigma)_{\rm f}}-1\right)\frac{V_{\rm i}}{V_{\rm i}^{\rm max}}$. The linear behavior expected from equation~(\ref{eq:sum2}) is indeed observed: we measure $K_{\rm f}=2.4~10^{-2}$  and $K_{\rm i}=3.9~10^{-4}$. The horizontal error bars are fairly large, since there is a non negligible  uncertainty in computing the actual volume of the impellers that plays a role in the induction effect. This is because the blades are indeed attached with stainless steel screws that may prevent the optimal development of electrical currents in more conducting materials.\\

The inhomogeneity of magnetic diffusivity at the impellers is also expected to produce an added induction in the case of an axially applied field -- in this case, however, a velocity shear at the fluid-impeller interface is necessary (otherwise no variation of flux can cause induction). Using again the calculation by Herzenberg and Lowes~\cite{Herzenberg:PTRS:57} as a reference, we expect a variation of the azimuthal induced field with disks and blades materials. Figure~\ref{fig:slopes} (red triangles) shows that it is indeed the case, but the evolution is much shallower. The explanation lies in the fact that the effect is localized in the vicinity of the impellers -- it decays as the inverse third power of the distance to the disk~\cite{Herzenberg:PTRS:57}(equation 3.21)) and, in this case, there is no global effect of the boundary condition that constrains the currents and help ensure that a sizable magnetic field can be detected in the mid-plane of the flow (where the measurement probes are located). \\

We thus find that our modelization, which attributes the induction to conventional effects from the shearing motion of the fluid plus an additional contribution due to the inhomogeneity of the magnetic diffusivity of the impellers compared to that of the fluid, gives an adequate interpretation of our experimental data.

\subsection{Illustration: induction effects linked to helicity}\label{sec:exp_alpha}
When the fluid is set into motion by the rotation of a single impeller, the mean flow has a strong helical component. It results from the rotational entrainment and axial flow generated by the impeller acting as a centrifugal pump. When a transverse magnetic field is applied, this swirling motion induces an axial component through what has been termed the Parker effect~\cite{Petrelis:PRL:03}. Its evolution with the impeller rotation rate is quadratic because it invokes both the azimuthal and the axial velocity of the fluid. When impellers of high magnetic permeability are being used we expect an additional effect as described above, again being made accessible to measurements in the mid-plane because of the boundary condition at the vessel boundary. Moreover, this last effect is dominant at the lateral wall of the vessel, so we expect to observe: (i) induction characteristics corresponding to the Parker effect near the axis of the cylinder (it varies quadratically with the impeller rotation rate); (ii) induction characteristics associated to the solid rotation of the impeller in a transverse applied field, as described in the previous section (it varies linearly with the impeller rotation rate). 

\begin{figure}[h!]
\rightline{\includegraphics[width=0.85\columnwidth]{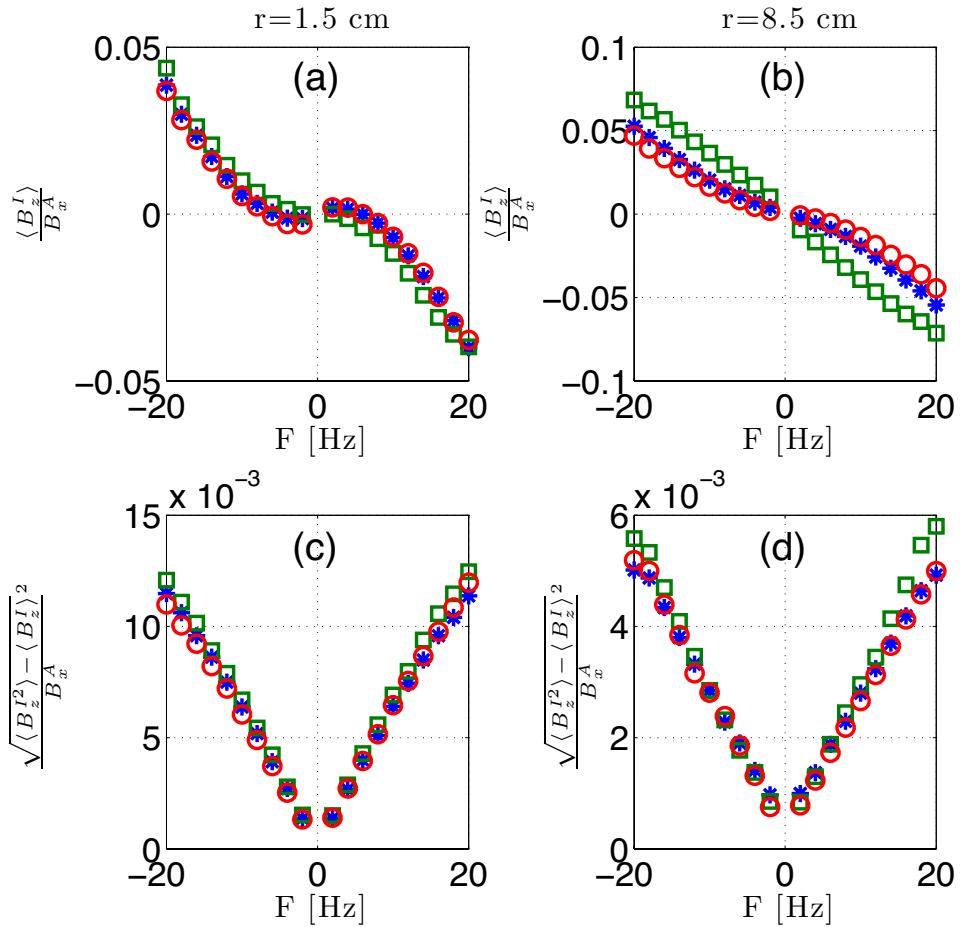}}
\caption{Evolution of the normalized time averaged induced field $\langle 
B^I_z\rangle$ as a function of the impellers rotation rate at radial position (a) $r/R = 
0.15$ and (b) $r/R = 0.87$. Corresponding $rms$ amplitudes in (c) and (d). Symbols: ($\star$) stainless steel impellers, ($\circ$) iron impellers,  ($\square$) soft-iron impellers. }
\label{fig:alpha}
\end{figure}

Our measurements correspond to ($F_2\neq 0 $ and $F_1 = 0$ in the configuration sketched in figure~\ref{fig:setup}(b)). The applied field is transverse, ${\mathbf B^A} = -B^A_x \hat{x}$. The evolution of the time-averaged induced field $\langle B^I_z \rangle$ measured in the mid-plane is shown in figure~\ref{fig:alpha}(a,b) for the most inner probe in the array ($r/R = 0.15$) and the most outer probe ($r/R = 0.87$). As can be seen, in the center of the flow, the induction varies quadratically with the impeller rotation rate, in a manner which is independent of its material. On the other hand, nearer the outer wall, the evolution, initially quadratic, becomes linear when soft iron impellers are used. In all cases, the fluctuations shown in figure~\ref{fig:alpha}(c,d) show no dependence with the impeller material. These features are consistent with the interpretation given in the previous section.

\section{Discussion and Conclusions}
Our observations show that the driving of von K\'arm\'an flows with ferromagnetic impellers have a significant impact on magnetic induction processes. The response of the flow to an externally applied magnetic field reveals additional contributions to the induced field. Moreover, at the intermediate magnetic Reynolds numbers probed with our gallium flow, the additional contributions vary linearly with the change in $\mu\sigma$ between fluid and impeller. Finally, as this additional contribution is associated with the (controlled) motion of the impellers, it has much less fluctuations than the induction originating solely from the flow velocity gradients.\\

Such effects may have a significant impact on dynamo processes in laboratory experiments. They could be quite important in the generation of the VKS dynamo~\cite{Monchaux:PRL:07}. Expanding on our observations, one may expect that the combination of velocity shear and  magnetic diffusivity discontinuity at the impellers generates a quite strong $\omega$-effect in the vicinity of the soft iron impellers. As a result, an axial magnetic field would be efficiently converted into a toroidal field in this region. For the regeneration of the axial field, several types of $\alpha$-effects have been proposed~\cite{Monchaux:PoF:09}. In the above scenario, the VKS dynamo generates an axial dipole from  $\alpha-\omega$ processes which are both localized in the vicinity of the ferromagnetic impellers, which also has the following major contributions: \\
- their large $\mu\sigma$ value and associated enhancement of induction effects (as measured in section~\ref{sec:transverse}) helps bring a dynamo threshold within the range of $R_m$ values accessible to the experiment,\\
- they promote an axial magnetic field. It would help understand why the actual dynamo field generated with soft iron impellers is dominated by an axial dipole, in contrast with numerical simulations made for homogeneous magnetic conditions which predicted a transverse dipole,\\
- the localization of dynamo sources near the impellers may help understand why the evolution of the VKS dynamo field shares many features with low dimensional chaos~\cite{Ravelet:PRL:08}, as would result from non-linear interactions of two weakly coupled dynamos. \\

Further measurements will of course be needed. They involve the precise measurement of the magnetic fields in the very vicinity of the impellers -- an endeavor that requires major changes in the experimental setup and involves several technical challenges. Detailed induction measurements in the sodium (VKS) experiment would also be needed to clarify the evolution with the magnetic Reynolds, as the current gallium studies are restricted to $R_m$ values of order unity. \\

\noindent{\bf Acknowledgements}\\
The assistance of Marc Moulin is gratefully acknowledged. This work is funded under contract ANR-08-BLAN-0039-01. \\

\bibliographystyle{unsrt}
\bibliography{MHD}

\begin{thebibliography}{10}

\bibitem{Kageyama:PP:95}
A.~Kageyama, T.~Sato, K.~Watanabe, R.~Horiuchi, T.~Hayashi, Y.~Todo, T.~H.
  Watanabe, and H.~Takamaru.
\newblock Computer simulation of a magnetohydrodynamic dynamo 2.
\newblock {\em Physics of Plasmas}, {2}({5}):{1421--1431}, {1995}.

\bibitem{Muller:01}
U.~M\"uller and L.~B\"uhler.
\newblock {\em Magnetofluiddynamics in Channels and Containers}.
\newblock Springer (Berlin, Heidelberg), 2001.

\bibitem{Glatzmaier:PEPI:95}
G.A. Glatzmaier and P.H. Roberts.
\newblock A 3-dimensional convective dynamo solution with rotating and finitely
  conducting inner-core and mantle.
\newblock {\em Physics of the Earth and Planetary Interiors},
  {91}({1-3}):{63--75}, {1995}.

\bibitem{Glatzmaier:PhysD:96}
G.A. Glatzmaier and P.H. Roberts.
\newblock {An anelastic evolutionary geodynamo simulation driven by
  compositional and thermal convection}.
\newblock {\em {Physica D}}, {97}({1-3}):{81--94}, {1996}.

\bibitem{Stefani:ZAMM:08}
F.~Stefani, A.~Gailitis, and G.~Gerbeth.
\newblock {Magnetohydrodynamic experiments on cosmic magnetic fields}.
\newblock {\em Zeitschrift f\"ur Angewandte Mathematik und Mechanik},
  {88}({12}):{930--954}, {2008}.

\bibitem{Verhille:SSR:09}
G.~Verhille, N.~Plihon, M.~Bourgoin, P.~Odier, and J.-F. Pinton.
\newblock Laboratory dynamo experiments.
\newblock {\em Space Science Reviews}, 10.1007/s11214-009-9546-1, 2009.

\bibitem{Monchaux:PRL:07}
R.~Monchaux, M.~Berhanu, M.~Bourgoin, M.~Moulin, P.~Odier, J.-F. Pinton,
  R.~Volk, S.~Fauve, N.~Mordant, F.~P\'etr\'elis, A.~Chiffaudel, F.~Daviaud,
  B.~Dubrulle, C.~Gasquet, L.~Mari\'e, and F.~Ravelet.
\newblock {Generation of a magnetic field by dynamo action in a turbulent flow
  of liquid sodium.}
\newblock {\em {Phys.\ Rev.\ Lett}}, {98}({4}):{044502}, {2007}.

\bibitem{Monchaux:PoF:09}
R.~Monchaux, M.~Berhanu, S.~Auma\^itre, A.~Chiffaudel, F.~Daviaud, B.~Dubrulle,
  F.~Ravelet, S.~Fauve, F.~Mordant, N.and~P\'etr\'elis, M.~Bourgoin, P.~Odier,
  J.F. Pinton, N.~Plihon, and R.~Volk.
\newblock {The von K\'arm\'an Sodium experiment: Turbulent dynamical dynamos}.
\newblock {\em Phys. Fluids}, {21}({3}):{035108}, {2009}.

\bibitem{Marie:EPJB:03}
L.~Marie, J.~Burguete, F.~Daviaud, and J.~Leorat.
\newblock {Numerical study of homogeneous dynamo based on experimental von
  Karman type flows}.
\newblock {\em European Physical Journal B}, {33}({4}):{469--485}, {2003}.

\bibitem{Oxley:JMMM:09}
P.~Oxley, J.~Goodell, and R.~Molt.
\newblock {Magnetic properties of stainless steels at room and cryogenic
  temperatures}.
\newblock {\em Journal of Magnetism and Magnetic Materials},
  {321}({14}):{2107--2114}, {2009}.

\bibitem{FEMM42}
FEMM 4.2.
\newblock Finite elements method magnetics.
\newblock {\em Software, http://www.femm.info/}, page v4.2, 2009.

\bibitem{Handbook:94}
D.~Lide (ed.), editor.
\newblock {\em Handbook of chemistry and physics}.
\newblock CRC Press, 14th edition edition, 1994.

\bibitem{Daviaud04a}
F.~Daviaud L.~Mari\'e.
\newblock Experimental measurement of the scale-by-scale momentum transport
  budget in a turbulent shear flow.
\newblock {\em Physics of Fluids}, 16(2):457--461, 2004.

\bibitem{Odier:PRE:98}
P.~Odier, J.-F. Pinton, and S.~Fauve.
\newblock Advection of a magnetic field by a turbulent swirling flow.
\newblock {\em Phys. Rev. E}, 58:7397, 1998.

\bibitem{Bourgoin:MHD:04}
M.~Bourgoin, R.~Volk, P.~Frick, S.~Khripchenko, P.~Odier, and J.-F. Pinton.
\newblock Induction mechanisms in von {K}\'arm\'an swirling flows of liquid
  gallium.
\newblock {\em Magnetohydrodynamics}, 40:13, 2004.

\bibitem{Bourgoin:PoF:04}
M.~Bourgoin, P.~Odier, J.-F. Pinton, and Y.~Ricard.
\newblock An iterative study of time independent induction effects in
  magnetohydrodynamics.
\newblock {\em Physics of Fluids}, 16:2529, 2004.

\bibitem{ravelet2004}
F.~Ravelet, L.~Mari\'e, A.~Chiffaudel, and F.~Daviaud.
\newblock Multistability and memory effect in a highly turbulent flow:
  Experimental evidence for a global bifurcation.
\newblock {\em Phys. Rev. Lett.}, 93:164501, 2004.

\bibitem{Volk:PoF:06}
R.~Volk, P.~Odier, and J.-F. Pinton.
\newblock {Fluctuation of magnetic induction in von Karman swirling flows}.
\newblock {\em {Physics of Fluids}}, {18}({8}):{085105}, {2006}.

\bibitem{Herzenberg:PTRS:57}
A.~Herzenberg and F.J. Lowes.
\newblock Electromagnetic induction in rotating conductors.
\newblock {\em Phil. Trans. Roy. Soc. London A}, {249}({970}):{507--584},
  {1957}.

\bibitem{Weiss:1966}
N.O. Weiss.
\newblock The expulsion of magnetic flux by eddies.
\newblock {\em Proceedings of the Royal Society, A}, A293, 1966.

\bibitem{PrivateNoreGisecke}
C.~Nore and A.~Gisecke.
\newblock Influence of ferromagnetic disks on the vks dynamo.
\newblock {\em Private Communication}, 2009.

\bibitem{Petrelis:PRL:03}
F.~P\'etr\'elis, M.~Bourgoin, L.~Mari\'e, J.~Burguete, A.~Chiffaudel,
  F.~Daviaud, S.~Fauve, P.~Odier, and J.-F. Pinton.
\newblock Nonlinear magnetic induction by helical motion in a liquid sodium
  turbulent flow.
\newblock {\em Phys. Rev. Lett.}, 90:174501, 2003.

\bibitem{Ravelet:PRL:08}
F.~Ravelet, M.~Berhanu, R.~Monchaux, S.~Aumaitre, A.~Chiffaudel, F.~Daviaud,
  B.~Dubrulle, M.~Bourgoin, P.~Odier, N.~Plihon, J.-F. Pinton, R.~Volk,
  S.~Fauve, N.~Mordant, and F~P\'etr\'elis.
\newblock {Chaotic dynamos generated by a turbulent flow of liquid sodium.}
\newblock {\em { Phys. Rev. Lett. }}, {101}({7}):{ 074502}, {2008}.

\end{thebibliography}

\end{document}